\documentclass{emulateapj}
\usepackage{graphicx}
\usepackage{amsmath}
\usepackage{amsfonts}
\usepackage{amssymb}
\usepackage{natbib}
\usepackage{textcomp}
\usepackage{gensymb}
\usepackage{url}
\begin{document}

\title{Observations and modeling of the emerging EUV loops in the quiet Sun as seen with the Solar Dynamics Observatory}

\author{LP. Chitta$^{1,2}$, R. Kariyappa$^2$, A. A. van Ballegooijen$^1$, E. E. DeLuca$^1$, S. S. Hasan$^2$, and A. Hanslmeier$^3$}
\affil{$^1$Harvard-Smithsonian Center for Astrophysics, 60 Garden Street MS-15, Cambridge, MA 02138, USA}
\affil{$^2$Indian Institute of Astrophysics, Bangalore 560 034, India}
\affil{$^3$Institut f\"{u}r Physik, IGAM, Universit\"{a}t Graz, Universit\"{a}tsplatz 5, 8010 Graz, Austria}

\begin{abstract}
We used data from the Helioseismic and Magnetic Imager (HMI), and Atmospheric Imaging Assembly (AIA)
on the \textit{Solar Dynamics Observatory} (SDO) to study coronal loops at small scales, emerging in
the quiet Sun. With HMI line-of-sight magnetograms, we derive the integrated and unsigned
photospheric magnetic flux at the loop footpoints in the photosphere. These loops are bright in the
EUV channels of AIA. Using the six AIA EUV filters, we construct the differential emission measure
(DEM) in the temperature range $5.7 - 6.5$ in log $T$ (K) for several hours of observations. The
observed DEMs have a peak distribution around log $T \approx$ 6.3, falling rapidly at higher
temperatures. For log $T <$ 6.3, DEMs are comparable to their peak values within an order of
magnitude. The emission weighted temperature is calculated, and its time variations are compared
with those of magnetic flux. We present two possibilities for explaining the observed DEMs and
temperatures variations. (a) Assuming the observed loops are comprised of hundred thin strands with
certain radius and length, we tested three time-dependent heating models and compared the resulting
DEMs and temperatures with the observed quantities. This modeling used Enthalpy-based Thermal
Evolution of Loops (EBTEL), a zero-dimensional (0D) hydrodynamic code. The comparisons suggest that
a medium frequency heating model with a population of different heating amplitudes can roughly
reproduce the observations. (b) We also consider a loop model with steady heating and non-uniform
cross-section of the loop along its length, and find that this model can also reproduce the observed
DEMs, provided the loop expansion factor $\gamma \sim$ 5 - 10. More observational constraints are
required to better understand the nature of coronal heating in the short emerging loops on the quiet
Sun.
 \end{abstract}

\keywords{Sun: photosphere --- Sun: surface magnetism --- Sun: atmosphere --- Sun: corona}
\maketitle
\clearpage
\section{INTRODUCTION}
\label{sec:intro}
A part of the magnetic field originating in the photospheric sub-surface layers reaches higher up in
the solar atmosphere and forms loop like structures, the building blocks of solar corona. These
loops harbor plasma, which is heated up to a few million Kelvin, much higher than the photospheric
temperature. Finding the source and nature of energy required to heat the corona along with the
process of heating is one of the most sought after questions in the field of astrophysics~\citep[for
reviews on coronal heating, see for
example][]{1993SoPh..148...43Z,1996SSRv...75..453N,2006SoPh..234...41K,lrsp-2010-5}. Studying the
dynamics of the plasma filled loops is important to understand the heating mechanisms responsible
for these high temperatures. Observational, theoretical, and numerical advances have been made over
several decades to understand the physics involved in these processes. Some of the early works on
this subject include the ideas of damping of magnetohydrodynamic waves in the lower corona to heat
the solar atmosphere~\citep[][for a recent review on waves in solar corona,
see~\citealt{2005LRSP....2....3N}]{1947MNRAS.107..211A}. 

From the early x-ray observations~\citep{1973SoPh...32...81V}, it became evident that the solar
corona is confined in the form of loops outlined by the underlying photospheric magnetic field. 
Later, \citet{1978ApJ...220..643R}, gave an analytical model for the quiescent coronal loops,
assuming that these structures are in hydrostatic equilibrium. They suggested that the observations
are indicative of a steady-state heating process.
\citet{1988ApJ...330..474P},~\citet{1994ApJ...422..381C}, and~\citet{1997ApJ...478..799C} put
forward the idea of intermittent and impulsive (nanoflare) heating, as a viable mechanism. It is now
generally believed, and widely accepted that the magnetic field plays an important role in
generating and transporting the energy required to maintain the temperatures of the corona. It
remains unclear and difficult to identify the dominant process responsible for heating of the solar
atmosphere.

As the diagnostics of tenuous coronal plasma improved with the advent of high spatial and temporal
resolution space based instruments, an alternate but relevant debate emerged within the community,
namely, the frequency of required heating events. The plasma filled in the loops respond to the
impulse of heating, and this depends on whether the plasma is reheated before it is completely
cooled down (high frequency model --- steady heating), or not (low frequency model --- nanoflares).
Should either of these models operate, they predict certain physical properties of the loops, which
can be compared with the observations~\citep[see][for a broad review on coronal loop observations
and modeling]{lrsp-2010-5}. 

With a wide range of field strengths and sizes of magnetic elements, coronal loops also have wide
temperature and length distributions. Usually the loops are classified as ``hot'' ($T > 2-3$ MK),
and ``warm''($T \approx 1-2$ MK) depending on their temperature regime. Both steady and impulsive
heating models have been extensively used to explain the observed temperatures, loop intensity
structure e.t.c. Studies indicate that the hot plasma is consistent with both steady heating
models~\citep{2010ApJ...711..228W,2011ApJ...740....2W}, and impulsive heating
models~\citep{2010ApJ...723..713T,2012ApJ...753...35V}. The warm loops are found to be continuously
evolving and not in equilibrium~\cite[see for example,][]{2009ApJ...695..642U}, and their properties
are well explained by impulsive heating models~\citep{2003ApJ...582..486S}. It is also suggested
that the age of an active region might play an important role in determining the dominance of one
process over the other~\citep{2012ApJ...756..126S,2012ApJ...761...21U}.

The active regions are well studied both in terms of observations and modeling. 
However, the situation is not so clear in the case of small loops in the quiet
Sun. The classification of ``hot'' and ``warm'' loops may not be relevant in
these features, owing to their compact magnetic structure and narrow
temperature range compared to the active regions. These short loops are
connected to magnetic bipoles in the photosphere. Their origin can be traced to
either flux emergence, or convergence of opposite polarities with reconnection.
The magnetic fluxes associated with these regions are typically in the range of
$10^{19} - 10^{20}$ Mx. The electron number density in such loops, measured
using density sensitive lines is in the order of $10^9$ cm$^{-3}$
~\citep{2005A&A...435.1169U,2008A&A...492..575P,2010ApJ...710.1806D}.

In this study, we are primarily interested in understanding the nature of the heating that produces
the observed $1-2$ MK temperature in these small bipoles, in particular the frequency of heating
events. Also, to better understand the relation between photospheric magnetic field and the coronal
loop temperatures, we chose to study emerging flux events. In these events it is easy to identify
the loops, and their footpoints in the photosphere. We follow their formation and evolution over many hours. In the
following section, we present the observational results. Section~\ref{sec:model} describes the loop
modeling and the simple heating models we tested in this work. Finally, we summarize the results,
and discuss some relevant aspects that require further investigation.

\section{OBSERVATIONAL RESULTS}
\label{sec:obsvr}
In this section we give a brief note on the datasets used, and present the results derived, namely,
the photospheric magnetic flux, and coronal temperatures. The line-of-sight magnetograms observed
with the Helioseismic and Magnetic Imager~\citep[HMI,][]{2012SoPh..275..207S,2012SoPh..275..229S}, and the intensity
images from the EUV channels of Atmospheric Imaging Assembly~\citep[AIA,][]{2012SoPh..275...17L} are
used. HMI and AIA are two of the three instruments onboard \textit{Solar Dynamics
Observatory}~\citep[SDO,][]{2012SoPh..275....3P}. Data are taken from 2011 February 10, and 2012
March 17 observations, spanning for about 12 hr each. SDO observes the full disk of the Sun
continuously in different filters with a high cadence of 12 s. We selected a region near disk center
with the criteria that, we see emerging magnetic field and coronal loops close to the beginning of
the selected time sequence. A few cases of evolved bipoles are also considered.

AIA data contain time sequences from 94~\AA, 131~\AA, 171~\AA, 193~\AA, 211~\AA, and 335~\AA~EUV
channels. Data are processed with standard procedures available in the \textit{solarsoft} library.
Alignment between the data from all these channels is crucial. Using 171~\AA~images as reference,
and cross-correlation technique, we aligned all data to within a pixel. The emerging bipoles are
identified both in HMI, and AIA. The tracked data cubes of such bipoles are extracted for further
analysis.  To enhance the signal-to-noise ratio, we prepare the 12 s cadence AIA data to 1 minute
cadence by averaging five exposures in each channel. Next, to derive the physical properties of the
plasma, we adopt the differential emission measure (DEM), which is related to the electron number
density ($n_e$), and the line-of-sight plasma temperature gradient, and defined as
\begin{equation}
\varphi(T)=n^2_e\frac{dh}{dT}\label{demeq}.
\end{equation}

We use data from six AIA EUV channels, along with the filter responses\footnote{The filter
responses of 94 and 131~\AA~channels are empirically modified to include contributions from Fe {\sc
IX} and Fe {\sc XII} for 94~\AA, and from Fe {\sc VIII} and Fe {\sc XI} for 131~\AA. The revised
response functions can be obtained using \texttt{aia\_get\_response}~with a keyword
\texttt{chiantifix}, available in \textit{solarsoft}.} as input to construct DEM($T$)
(cm$^{-5}$~K$^{-1}$), at each pixel, using \verb+xrt_dem_iterative2.pro+~\citep[][distributed in
\textit{solarsoft}]{2004ASPC..325..217G,2004IAUS..223..321W}. In this program, initial DEM is
guessed and folded through the filter responses to generate model observations, which are
iteratively used to reduce the $\chi^2$ between the original and modeled observations. This program
uses a much tested IDL routine {\verb+mpfit.pro+}~\citep{2009ASPC..411..251M}, that performs a
Levenberg-Marquardt technique to solve the least-squares problem. 

\begin{figure}
\includegraphics[width=0.45\textwidth]{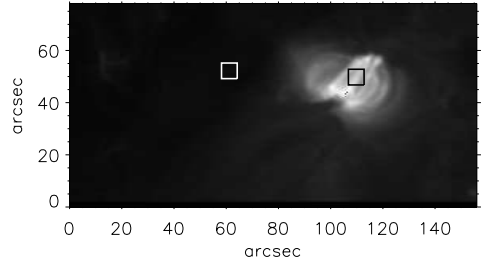}
\caption{A context image showing one of the analyzed bipoles as seen in AIA 193~\AA~channel. The
regions marked with black (near loop top), and white (background region) boxes are used for further
analysis to produce DEMs (see Figure~\ref{4dems}(a).) \label{context}}
\end{figure}

\begin{figure}
\includegraphics[width=0.45\textwidth]{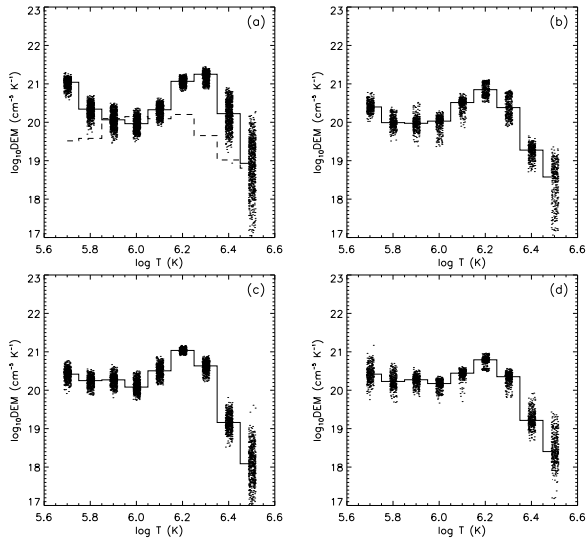}
\caption{DEMs of four bipoles obtained from the observations plotted as a function of temperature
for all times (dots). A small offset in temperature is given to the DEMs to show the temporal
distribution. Solid line is the temporal median of DEMs obtained for the respective cases. Observed
DEMs have a peak around log $T$ of $6.2 - 6.3$. Dashed histogram in panel (a) is the temporal median
of five DEMs, at random times, obtained from a \textit{background} region close to the corresponding
bipole. \label{4dems}}
\end{figure}

We show results from four emerging bipoles in this work. In Figure~\ref{context} we plot the
region-of-interest for one of the bipoles analyzed, as seen in AIA 193~\AA. The image saturates at
750 DN s$^{-1}$. For these bipoles, a series of DEMs are constructed near the loop top (for example
from a region with in the black box shown in Figure~\ref{context}) at each pixel in a $6'' \times
6''$ region, over several hours of observations. The predicted intensities from forward modeling of
the derived DEMs match the observed intensities with in the limits of errors. Since we restrict the
DEMs within a limited range of temperature, the predicted intensities will be lower limits of the
observed values. In Figure~\ref{4dems} we plot the average emission from this area as a function of
temperature (log $T$), for all times. Each panel corresponds to a bipole. The dots denote the time
dependence. At any given temperature, to show the emission distribution in time, we gave a small
offset to DEMs in temperature (and that is the reason we see a small spread of DEMs along log $T$).
Additionally, the temporal distributions also give a sense for the errors in the DEMs. The solid
lines are respective temporal medians for all DEMs. They have a peak close to log $T$ (K) of
$6.2-6.3$. At higher $T$, they show a rapid decline and also the DEMs are not well constrained. On
the other hand, at lower $T$, the emission stays comparable to the peak emission. Similar results
were obtained using \textit{Hinode}/EIS observations, but for a coronal hole bright
point~\citep[c.f. Fig. 12,][]{2010ApJ...710.1806D}.

We also note that there is no background subtraction to the data in DEM analysis. The small loop
structures we analyzed have their loop apex, and footpoints in the same plane along the
line-of-sight, much of the emission contribution may be primarily dominated by the loop apex with a
part of it originating from the footpoints. To compare the background contribution to the
resulting DEMs, for example (a), we considered a $10 \times 10$ pixel \textit{background} region
adjacent to that bipole (marked with a white box in Figure~\ref{context}). The DEMs are constructed
for this region at five random times, and the temporal median is plotted as a dashed histogram in
Figure~\ref{4dems}(a). This shows that the observed DEM lies well above the background, not only for
$\log T$ in the range $6.2 - 6.4$, but also at low temperatures ($\log T \le 5.8$). The small
scatter in the DEM at low $T$ suggest that this emission is real and is not an artifact of the DEM
inversion. The DEMs are now used to derive an emission weighted temperature ($T_{DEM}$)
using the relation
\begin{equation}
T_{DEM}=\frac{\Sigma\left\{\mbox{DEM}(T_i)T_i\Delta T_i \right\}}{\Sigma\left\{\mbox{DEM}(T_i)\Delta T_i\right\}}\label{emwt},
\end{equation}
where $\Delta T_i=0.1$ in log $(T)$, is the width of temperature bin around $T_i$. The time
variations of $T_{DEM}$ will be used later for comparison with loop models.

\begin{figure}
\includegraphics[width=0.45\textwidth]{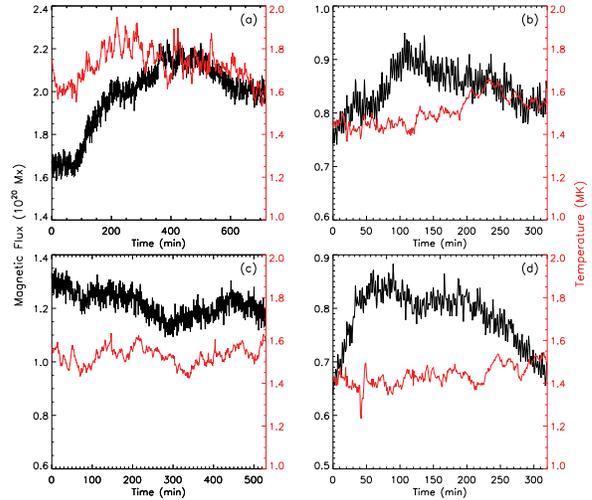}
\caption{Magnetic flux and loop temperatures for a sample of four bipoles. The black curves (left
axes) correspond to the integrated photospheric flux density of the bipoles. The red curves (right
axes) are the emission weighted, and averaged coronal temperature profiles for the respective
bipoles. Panels (a), (b), and (d) are for emerging bipoles, and panel (c) is for an evolved bipole.
\label{4examp}}
\end{figure}

Integrated unsigned magnetic flux of both polarities associated with these examples, as a function
of time, are also calculated from HMI\footnote{The two polarities are separated by a distance of
approximately 10 - 15 Mm in the photosphere.}. Such profiles of temperature and magnetic field for a
sample of four bipoles are plotted in Figure~\ref{4examp}. Three cases of newly emerging bipoles
(panels (a), (b), and (d)), and a case of an emerged bipole (panel (c)) are shown. The black curves
are time profiles of magnetic flux ($10^{19} - 10^{20}$ Mx), and red curves are temperatures in the
range of $1 - 2$ MK.

Though they all fall in a category of emerging/ emerged loops, there is no clear relation between
the magnetic flux at photosphere and the coronal loop temperature. In other words, it is not trivial
to directly relate the field changes in photosphere to the temperature fluctuations in the corona.
For example in Figure~\ref{4examp}(a), there is a strong correlation between the two physical
quantities in the long term trend, but in panel (b) the temperature seem to increase while the flux
decreases. In panels (c)-(d), it is more complicated. We suggest that every emerging bipole may
behave differently owing to its surrounding structures both in corona and photosphere. However, a
common signature is that the temperature fluctuates/ rises at some stage in the emergence process. 

\begin{figure}
\includegraphics[width=0.45\textwidth]{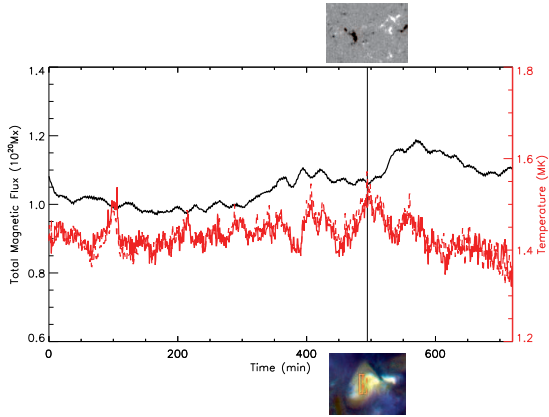}
\caption{Same as in Figure~\ref{4examp}. Integrated flux density (black curve), and the
temperatures (red solid and dashed curves) of an evolved bipole are plotted.  A snapshot of the
photospheric field configuration (top image, from HMI, $53''\times 39''$), and the corresponding
coronal loop structure (bottom composite image, from AIA 171~\AA, 193~\AA, 211~\AA~ channels,
$60''\times 48''$) are shown for a particular time as demarcated by the thin vertical line. The
solid and dashed red curves are the emission weighted temperatures derived from $4 \times 25$ pixel
($2''.4 \times 15''$) rectangular boxes, from the regions marked in the bottom image respectively. A
complete observed evolution of this example is presented as an animation, accompanying this
figure.\label{evolve}}
\end{figure}

To further illustrate this behavior, we consider another example of an emerged bipole. In
Figure~\ref{evolve} we plot the magnetic flux (black curve) and temperatures (red curves) this
bipole. A sample image of this example from a particular time is also shown above (magnetic
structure), and below (coronal loop) the plot. The thin vertical line demarcates the time of
snapshot. The solid and dashed red curves are average temperatures derived from two adjacent regions
(marked with solid and dashed lines in the image below the plot), of $2''.4 \times 15''$ size each.
In the accompanying animation, it is observed that, the drop in temperature after 500 minute is due
to the reconnection (in the corona) of the parent bipole with the adjacent opposite polarity
regions, changing the topology of the field,  and completely disrupting the main loop. Hence the
observed temperature of the loops originating from small ephemeral regions possibly depends on
various factors.  


\section{LOOP MODELING}
\label{sec:model}
Temperature profile of the loop is a good diagnostic for the loop dynamics but to get a better
picture, we also need to estimate the heating rate required to produce the observed temperatures. To
this end, we use Enthalpy-based Thermal Evolution of Loops~\citep{2008ApJ...682.1351K,
2012ApJ...752..161C}. EBTEL is a time dependent zero-dimensional (0D), hydrodynamic coronal loop
model. For a given loop half-length and volumetric heating rate, the code returns the loop
properties in terms of average temperature, density, and pressure of the loop and also the values of
these quantities at the loop apex (see the Appendix~\ref{sec:appendixA}).

\begin{figure}
\includegraphics[width=0.45\textwidth]{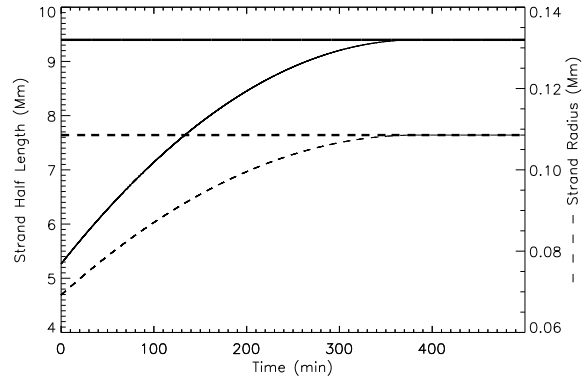}
\caption{Physical dimensions of the strands used in this study. Thick solid and dashed lines,
respectively, are the half length and radius of a constant strand. Thin solid and dashed curves,
respectively, are the half length and radius of an expanding strand. \label{volume}}
\end{figure}

We use EBTEL to model and derive the properties similar to the observed loops (we consider the
example shown in Figure~\ref{4examp}(a) for this purpose). The properties include the DEM as a
function of temperature, and the emission weighted temperature. We compare three different heating
scenario and discuss the results. For the models presented in next three subsections we make the
following assumptions: (a) A loop is comprised of hundred individual strands, each with a constant
length\footnote{This is only a rough estimate of the length based on the footpoint separation in the
photosphere.} of about 18 Mm, and a uniform radius of about 0.1 Mm. In Figure~\ref{volume} we plot
the length and radius of a single strand as a function of time (thick solid and dashed lines,
respectively). (b) Each strand is randomly heated with a certain heating profile over a period of
500 minutes. (c) The average values of various physical quantities over all the strands, represent
the properties of the whole loop. Along with these assumptions, the heating events are chosen such
that the modeled emission weighted temperatures closely match the observed temperatures.

In the Section~\ref{sec:case1} we describe the medium frequency heating model.
Section~\ref{sec:case2} deals with the low frequency heating model.  A medium frequency hybrid
heating model is discussed in the Sections~\ref{sec:case3a}, and~\ref{sec:case3b}. In
Section~\ref{sec:non-uni} we present an alternate explanation for the observed DEMs by considering
a non-uniform cross-section of the loop.

\subsection{MEDIUM FREQUENCY HEATING MODEL}
\label{sec:case1}
In the medium frequency heating model (case 1), individual strands are randomly heated with heating
rates having 50 - 100 s temporal fluctuations. These rates are generated by a sequence of random
numbers, and further filtering the signal within the desired band of periods. The base or minimum
heating rate is 10$^{-6}$ erg cm$^{-3}$ s$^{-1}$, and the amplitude of the fluctuations vary by up
to four orders of magnitude. The average heating rate for a single strand, over the entire duration
of 500 minutes is about 4~$\times$~10$^{-3}$ erg cm$^{-3}$ s$^{-1}$. 

\begin{figure}
\includegraphics[width=0.45\textwidth]{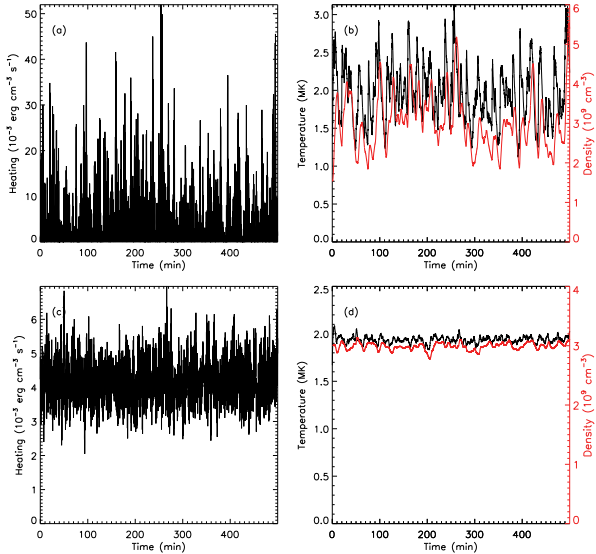}
\caption{Results from a medium frequency heating model (case 1). All the hundred strands are heated
with an approximately same average heating rate. (a) A representative input heating given to a
single strand with temporal fluctuations of 50 - 100 s. The base heating rate is 10$^{-6}$ erg
cm$^{-3}$ s$^{-1}$ for all the strands. The amplitudes of the heating rate fluctuate up to four
orders of magnitude. (b) The resulting temperature of the strand apex (black, left axis), and loop
density (red, right axis) for the heating profile shown in (a). (c) The average heating rate of
hundred random realizations. (d) The average loop apex temperature (black, left axis), and the
average loop density (red, right axis) averaged over hundred strands, representing an observed
loop.\label{case1}}
\end{figure}

A representative heating rate for one of the strands is plotted in Figure~\ref{case1}(a). The
plasma is reheated continuously before it is cooled to the equilibrium temperature due to base
heating. In panel (b) we plot the resulting temperature of the strand apex (black, left axis) along
with the resulting strand density (red, right axis). The temperature variations are $1-3$ MK within
in a single strand. Panel (c) is the average heating rate of all strands as a function of time. It
should be noted that the frequency of this average quantity is not a relevant factor in
distinguishing between various heating cases. Similarly in panel (d) we plot the average loop apex
temperature (black, left axis), and the average loop density (red, right axis). Since the observed
temperatures are derived from weighing the emission distribution, temperature of loop apex in panel
(d) cannot be directly compared with its observed counterpart. 

\begin{figure}
\includegraphics[width=0.45\textwidth]{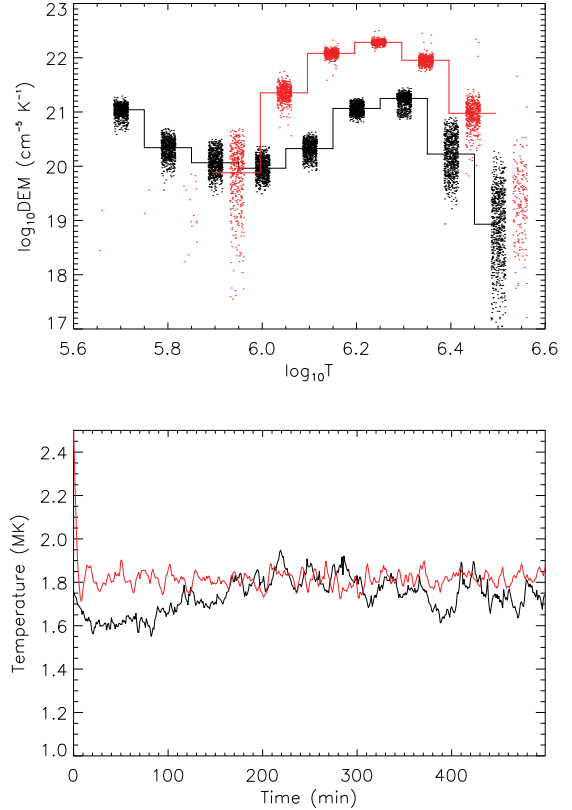}
\caption{Comparison of DEM results obtained for case 1 with the observations. Top panel: DEMs from
observations (black dots), and the modeled DEMs (red dots) are shown. All DEMs are given small
temperature offsets for a better visualization of the distributions. The black and red solid lines
are the temporal medians of observed and modeled DEMs respectively. Bottom panel: Emission weighted
temperature derived from observations (black) and modeling (red). \label{dem1}}
\end{figure}

In the top panel of Figure~\ref{dem1} we plot the observed DEMs (black dots). The observed DEMs
have a broad distribution in temperature with a peak at log $T$ of $6.3$ and another peak at log $T$
of $5.7$ (same as Figure~\ref{4dems}(a)). The modeled DEMs, which have a narrow distribution, are
plotted as red dots with a similar temperature offset. Since there are heating events occurring
almost continuously compared to the cooling time of the strands, the loop has no time to cool down
completely and the temperature stays steady, with small fluctuations. Because of this reason, all
the emission comes from a narrow distribution of temperatures, which is reflected in the modeled
results. 

Bottom panel is the resulting emission weighted temperatures from observations (black) and modeling
(red). Note that the range, and level of fluctuations in the temperature match very well but,
modeled DEM has a peak at log $T$ of 6.25, and the predicted emission about this temperature is at
least an order of magnitude more than the observed values. Further, the model predicts a weak or no
emission at lower $T$. By increasing the magnitude of heating rate to match the temperature at which
the peak emission occurs, will inherently increase the emission, and also the weighted temperature
well beyond the observed $T$. 

\subsection{LOW FREQUENCY HEATING MODEL}
\label{sec:case2}
\begin{figure}
\includegraphics[width=0.45\textwidth]{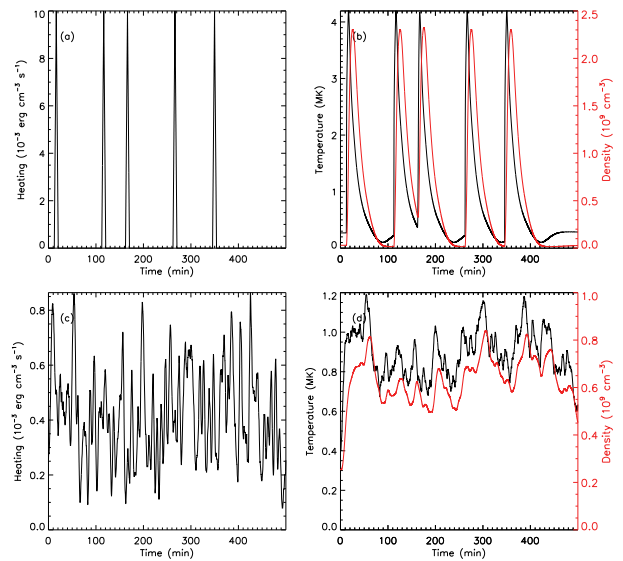}
\caption{Same as Figure~\ref{case1}. Results from a low frequency (impulsive) heating model (case
2). On an average, each strand is heated every 6000 s once, with a triangular heating pulse having a
maximum of 10$^{-2}$ erg cm$^{-3}$ s$^{-1}$, and a width of 500 s. \label{case2}}
\end{figure}

In the low frequency heating model (case 2), each strand is impulsively heated five times with an
average of 100 minutes interval between each impulse. Each triangular pulse has a width of 500 s and
a peak input of 10$^{-2}$ erg cm$^{-3}$ s$^{-1}$. Further, the base heating remains the same as in
case 1. In Figure~\ref{case2}(a) we plot a sample profile of heat input given to one of the strands.
Panel (b) is resulting temperature and density. Note that once the temperature reaches a maximum
value, it takes about 70 minutes for the strand to completely cool down.

\begin{figure}
\includegraphics[width=0.45\textwidth]{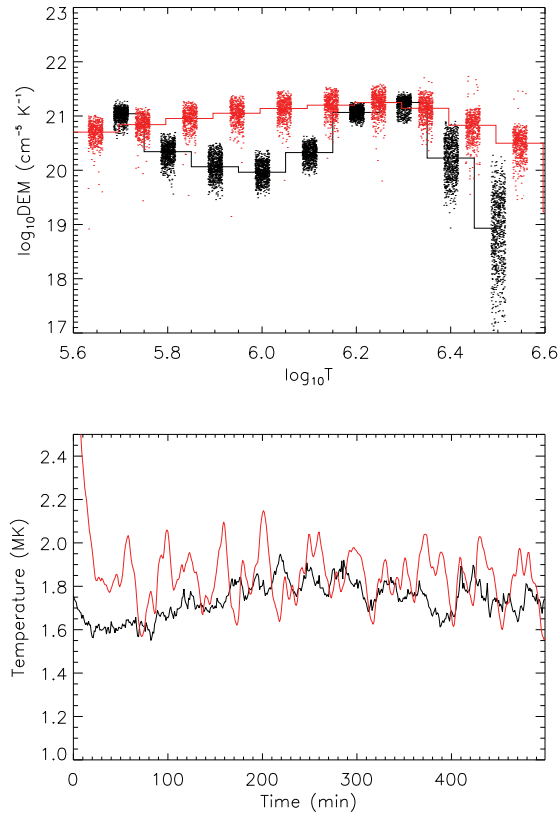}
\caption{Same as Figure~\ref{dem1}, but obtained for case 2.\label{dem2}}
\end{figure}

The average heating rate in panel (c) is less by a factor of 5 compared to that of case 1. It can
be seen that the temperature (panel (b)) in this model has a broad distribution, which is reflected
in a very broad DEM distribution shown in Figure~\ref{dem2} (top panel, red). At log $T$ of 6.5 the
model produces a well constrained DEM that is higher than the observed DEM, although the
observations are less constrained at those temperatures. Also, the model predicts an overall higher
emission at $\text{log}~(T)\approx6.0$

The predicted emission weighted temperature (bottom panel, red) is comparable with observed
temperature (bottom panel, black). The level of fluctuations and the short term trend in the red
curve are higher than what is seen in the observations. Furthermore, if the number of heating events
are fewer than what is considered here (five), but with stronger impulses, the fluctuations now
become noticeably large, and the observations should reveal these features. 

\subsection{MEDIUM FREQUENCY HYBRID HEATING MODEL}
\label{sec:case3a}
For cases 1 and 2 we adjusted the model parameters such that the DEM-weighted temperature
($T_{DEM}$) roughly matches the observed temperature for region 1. However, we find that the overall
structure and features of the predicted DEM($T$) does not match the DEM($T$) derived from AIA
observations. Therefore, neither of these models are fruitful in describing the $1-2$ MK emerging
loops in the quiet Sun. We suggest that the heating events may have a broad range, and/ or a
population of different heating amplitudes, influencing different strands. 

From the observational point of view, each strand in the loop is dynamically evolving and the
lifetime of this unit is not clearly known. New strands emerge with the photospheric flux and
replace the older ones in the loop. To investigate this problem further, a hybrid heating model
(case 3) has been considered. In this model, we start with a simple assumption that 20\% of the
strands are rapidly heated with an excess amount of average heating rate of fifty times more than
the remaining 80\% of the loops. All loops receive a base heating similar to that of cases 1 and 2. 

\begin{figure}
\includegraphics[width=0.45\textwidth]{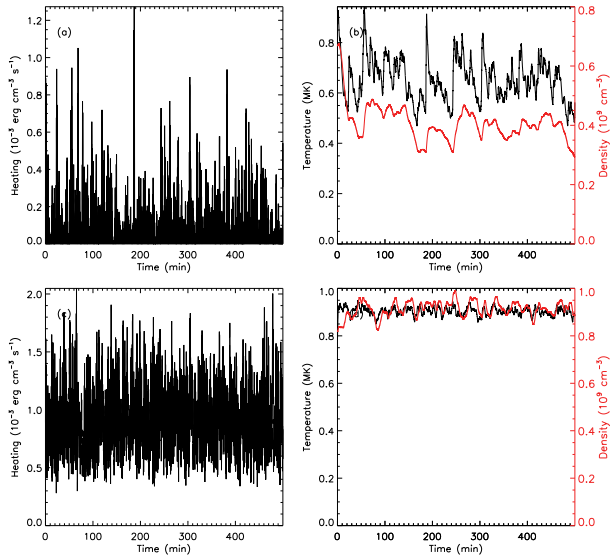}
\caption{Same as Figure~\ref{case1}. Results from a medium frequency hybrid heating model (case 3).
20\% of the strands are subjected to higher average heating inputs, but with same temporal
fluctuations as in case 1 (see text for details). Shown in panel (a) is an example of lower heating
rate case. \label{case3}}
\end{figure}

\begin{figure}
\includegraphics[width=0.45\textwidth]{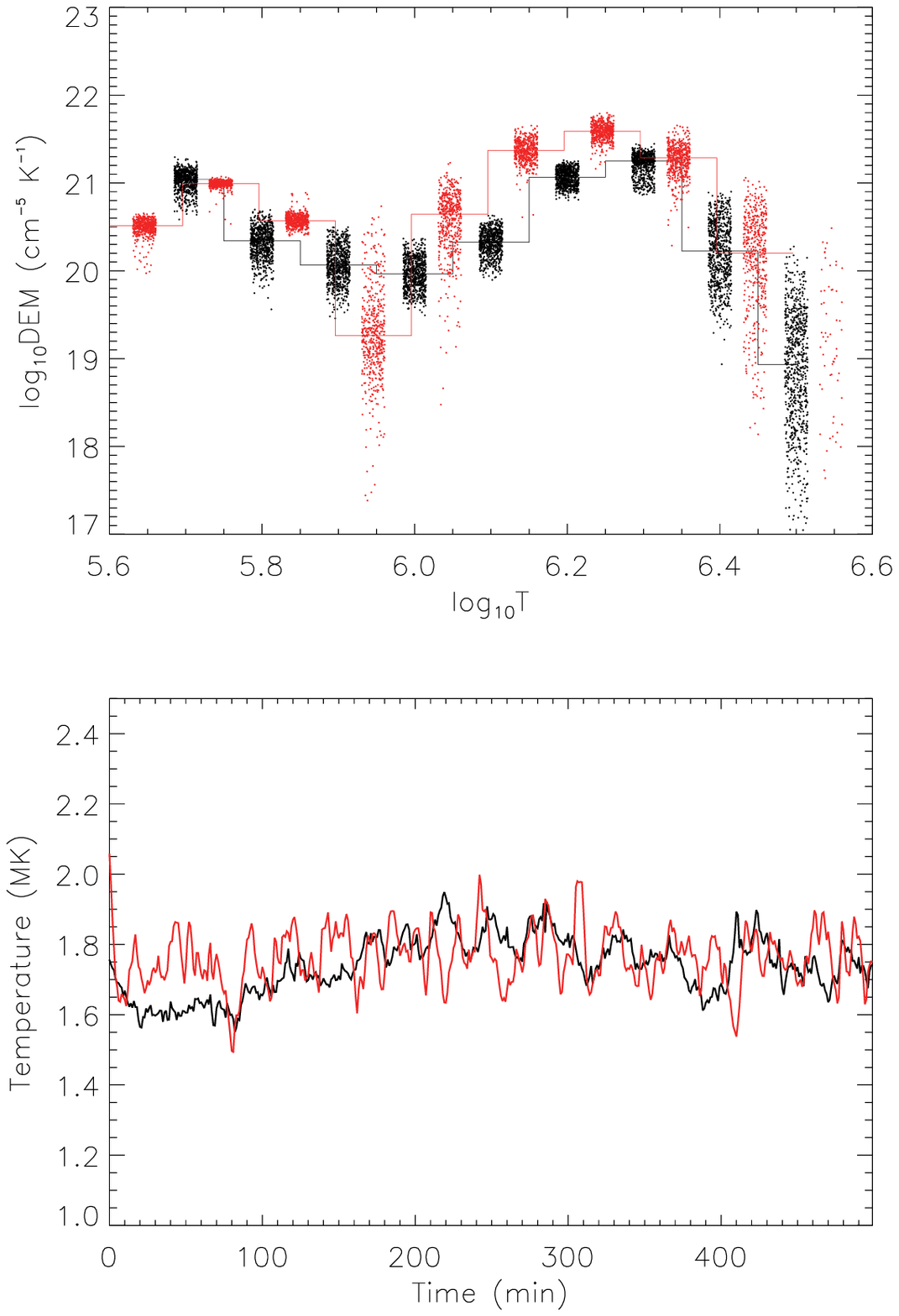}
\caption{Same as Figure~\ref{dem1}, but obtained for hybrid heating model (case 3).\label{dem3a}}
\end{figure}

In Figure~\ref{case3} we plot the heating rates, temperature, and densities and also the respective
average quantities. The profile in panel (a) is a low amplitude heating for a strand in the 80\%
population. The profiles shown in panel (b) are similar to that of Figure~\ref{case1}(b), except for
the overall lower values. Panels (c) and (d) show results averaged over all the strands, including
the 20\% that receive a higher level of heating. In Figure~\ref{dem3a} we show the DEM results for
this case. The top panel is for the observed (black) and predicted (red) DEMs. We see that the
predicted DEMs now have two distributions, clearly originating from the two populations of heating
events. It is interesting to note how closely the observed and predicted DEMs match. The emission
weighted temperature is shown in the bottom panel of Figure~\ref{dem3a}. The fairly well reproduced
quantities from this model suggest that a coronal loop, which has a bundle of many strands, can be
heated by considering different amplitudes of medium frequency heating events. This is certainly a
plausible assumption because, these emerging bipoles evolve continuously, and various reasons can
contribute to different heating episodes. 

\begin{figure}
\includegraphics[width=0.45\textwidth]{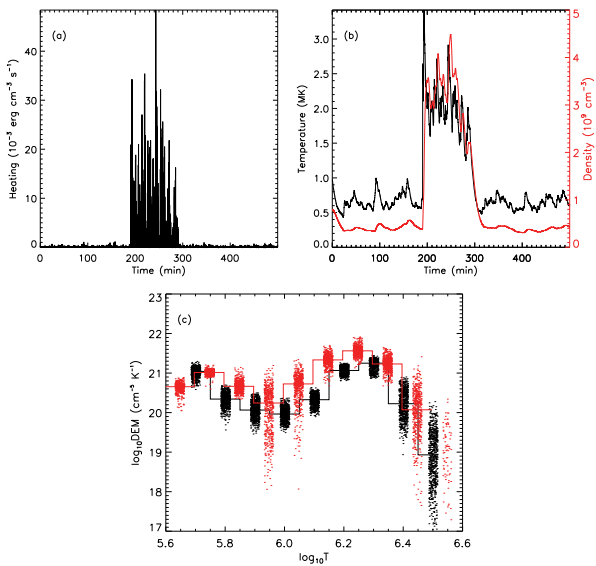}
\caption{Alternative version of case 3 in which, each strand receives high heating events 20\% of
its time, and the remaining time, low heating events. In panel (a), a sample heating profile is
plotted. The panels (b) shows strand apex temperature (black, left axis) and strand density (red,
right). The panel (c) is same as the top panel of Figure~\ref{dem3a}.\label{dem3aa}}
\end{figure}

Alternatively, we can also assume that each strand spends 20\% of its time being heated to higher
values (similar to the 20\% strands case described previously), and the remaining time to lower
values. Both the scenarios produce similar results. In Figure~\ref{dem3aa} we plot the results from
this alternative case. The panel (a) is a sample heating profile showing both low and high heating
events. In panel (b) the loop apex temperature and density are shown. In panel (c) the observed
(black), and predicted (red) DEMs are plotted along with their respective temporal medians. 

\subsection{A CASE OF EXPANDING LOOP}
\label{sec:case3b}
In general, the coronal loop length increases with time as it emerges through the solar atmosphere.
Also, the area as a whole, as the strength of the magnetic field drops with height, the area of
strand increases with time. Due to this expansion, filled in plasma may experience additional
adiabatic cooling effects, as the loop pressure and density are modified by the volume change (see
the Appendix~\ref{sec:appendixA}). We tested heating model described in case 3 on a slowly expanding
loop, comprised of hundred strands as explained in the previous sections.

\begin{figure}
\includegraphics[width=0.45\textwidth]{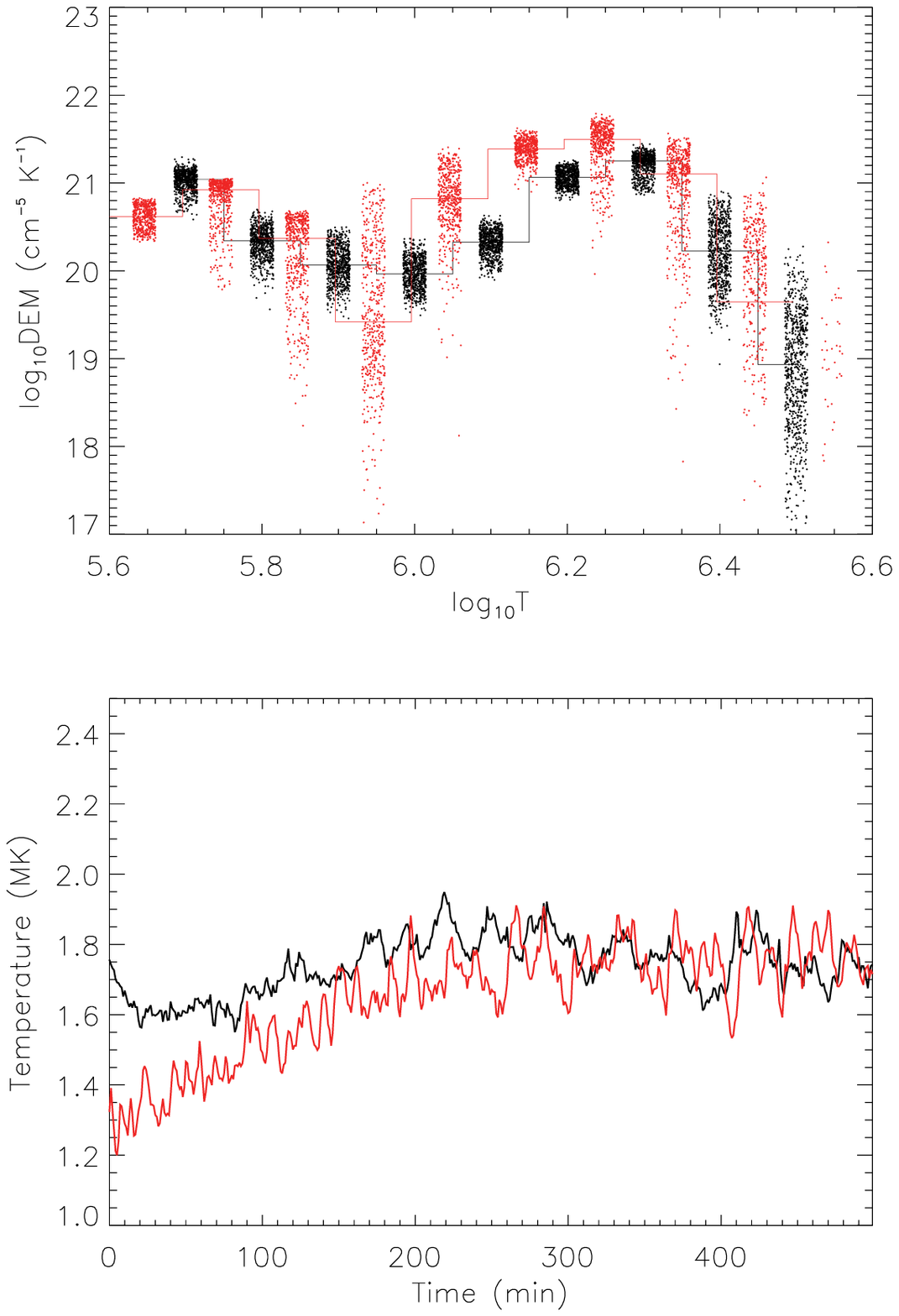}
\caption{Hybrid heating model with expanding length and radius as shown in Figure~\ref{volume}
(thin solid and dashed curves respectively). Both the length and radius of the loop vary slowly with
time. In this case the temperature (red, lower panel) increases as the length of the loop
increases.\label{dem3b}}
\end{figure}

The half length and radius of a single strand are shown as thin solid, and dashed curves in
Figure~\ref{volume}. The DEMs and $T$ are plotted in Figure~\ref{dem3b}. The way we consider the
volume expansion is that, each strand slowly expands for about 350 minutes and then the expansion
saturates to a constant value. This constant value matches with the length, and radius of the strand
chosen in all the cases.  It is observed that for a slowly expanding loop, the adiabatic cooling
effect can be negligible. There are two competing effects here. Under equilibrium conditions, the
temperature of the loop increases with the length. In our slowly expanding loop, the cooling is
compensated for with the length increase. But in reality, the rate of volume expansion can be
entirely different, and more rapid than what we considered here. These effects become important when
changes in the loop pressure and density due to expansion alone, and heating are comparable. 

\subsection{EFFECTS OF NON-UNIFORM CROSS-SECTION OF THE LOOP}
\label{sec:non-uni}
In the above EBTEL-based models the emission is assumed to come from the coronal portion of the
loop. EBTEL also predicts the emission from the transition region (TR) to model the lower
temperatures. However, the predicted TR emission is strong and rather flat relative to the
corona. Inclusion of up to $5-10\%$ of the TR emission will not affect the results, but adding more
contribution from the TR requires stronger heating to match the observed emission weighted
temperature. This results in a strong emission from the higher temperature, which is not observed.

\begin{figure*}%
\begin{minipage}{\textwidth}%
\begin{center}
\includegraphics[width=0.45\textwidth]{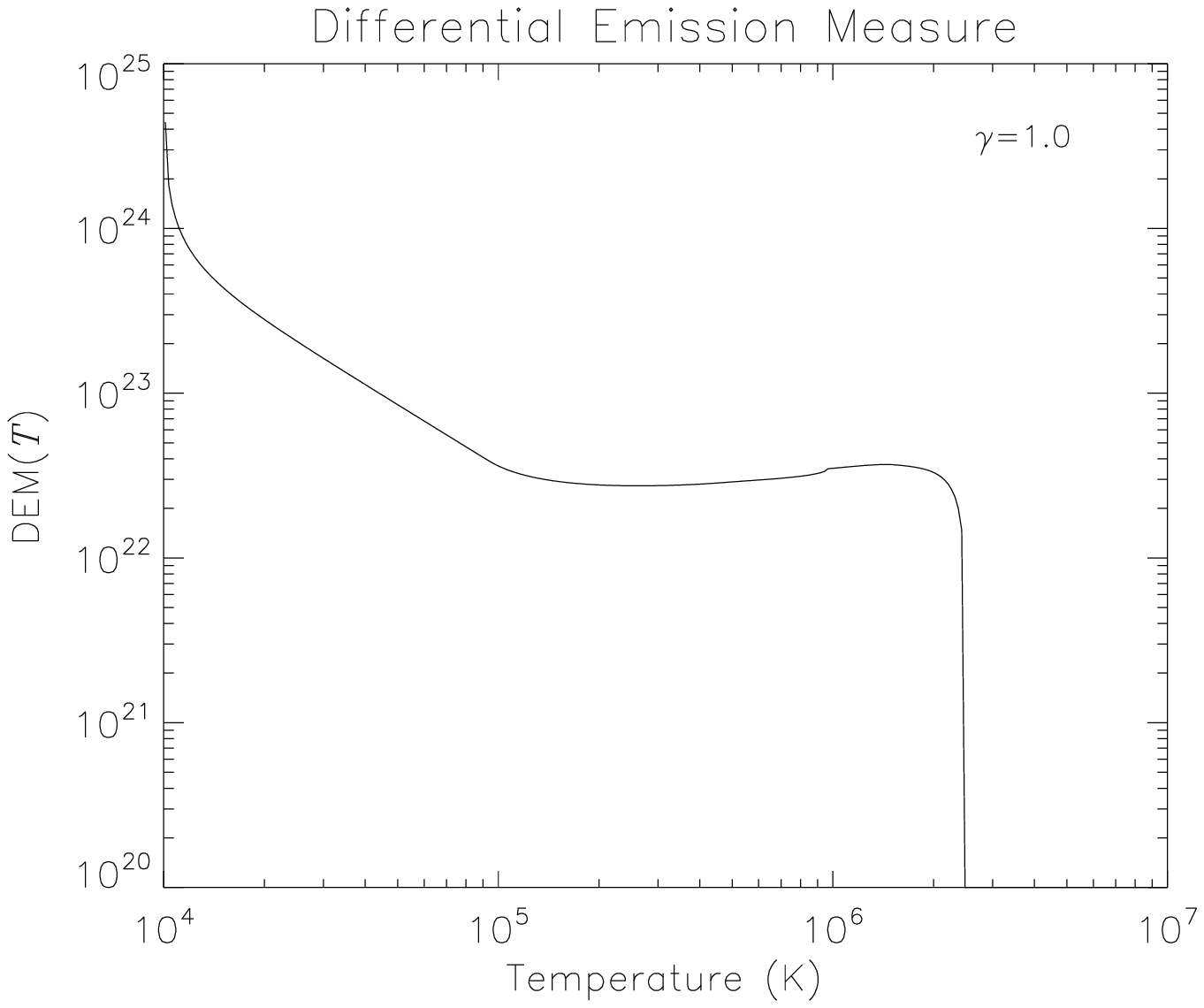}
\includegraphics[width=0.45\textwidth]{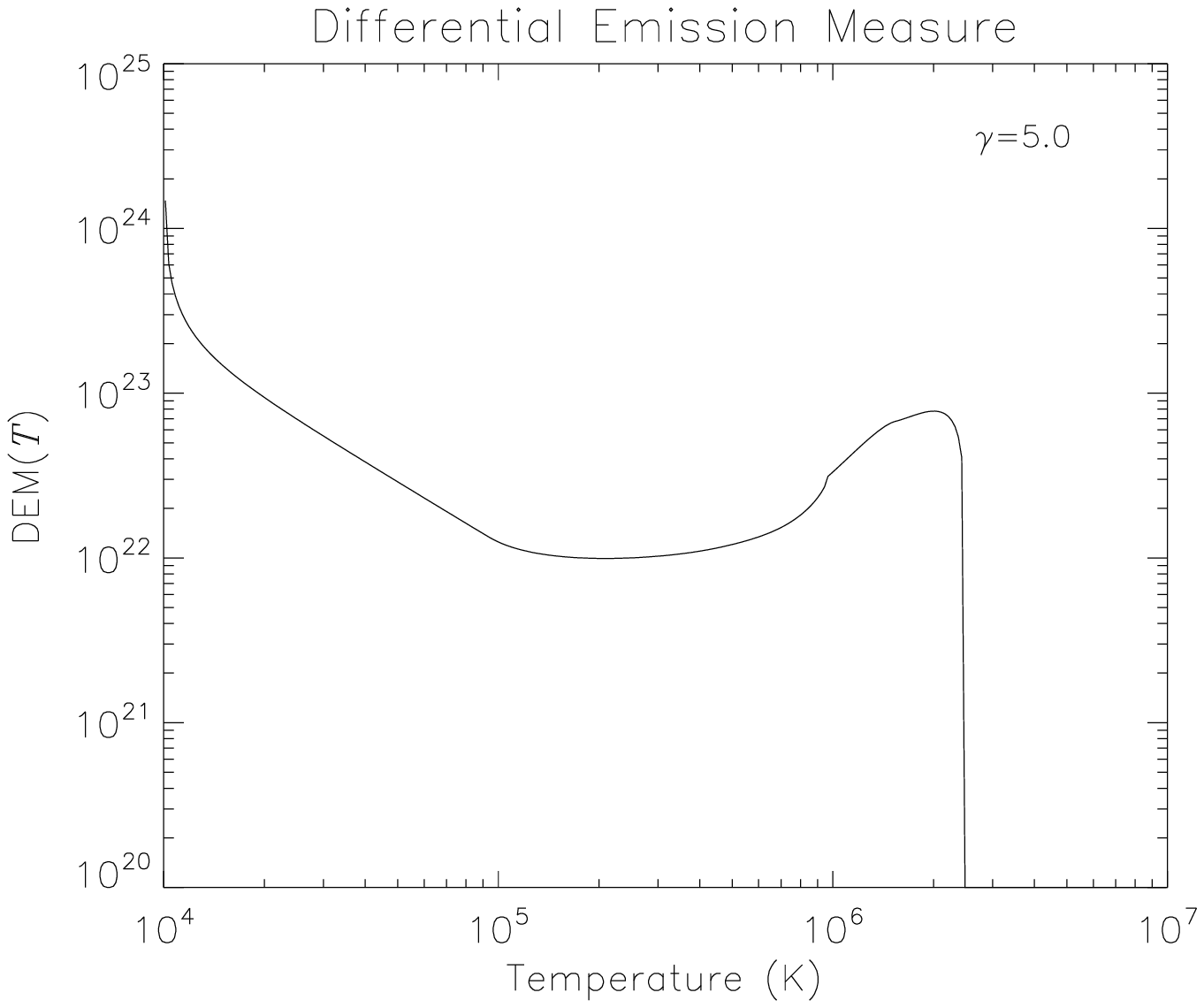}
\end{center}
\end{minipage}%
\caption{Results from a simple loop model assuming steady heating. Left panel: DEM of a loop with
expansion factor ($\gamma$) = 1. Note the flat DEM in the range of $10^5 < T $(K)$ <
10^6$. Right panel: Similar to the left panel, but with $\gamma = 5$. The DEM now shows a clear peak
at $T \approx 2$ MK.}%
\label{expan}%
\end{figure*}

One possible reason for the strong TR DEMs produced by EBTEL is that the model assumes a constant
cross-section over the length of the loop, whereas the loops on the Sun have significant expansion
factors ($\gamma$) between the loop footpoints in the TR and the loop top in the corona. Potential
field modeling of active regions~\citep[e.g.,][]{2012ApJ...746...81A} indicates $\gamma = 3 - 30$,
depending on height, and similar expansion factors may occur on the quiet Sun. When the
cross-sectional area $A$ of a loop increases with height, the volume of plasma at coronal
temperatures is increased relative to that at TR temperatures, so the slope of the DEM($T$) curve
becomes steeper and more consistent with observations.

To demonstrate this effect, we developed a simple loop model for the case that the cross-section $A$
varies along the loop. The heating is assumed to be steady in time. The model is described in
Appendix~\ref{sec:appendixB}. It allows us to compute the DEM($T$) for a single loop with a given
expansion factor $\gamma$, half-length $L$, and peak temperature $T_{\rm max}$ (we use $L = 9$ Mm).
We repeat the calculation for different peak temperatures ($6.0 < \log_{10} T_{\rm max} < 6.4$) and
compute the average DEM($T$). In Figure~\ref{expan} we plot the DEM results for $\gamma=1$ (left
panel) and $\gamma=5$ (right panel). Note that for a loop with uniform cross-section ($\gamma=1$),
DEM($T$) is flat for $T> 10^5$ K, similar to the DEMs predicted with the EBTEL
code~\cite[see][]{2008ApJ...682.1351K}. In contrast, for $\gamma=5$ the peak value of the DEM in the
corona is about 8 times its value in the TR, similar to the observed DEMs (see Figure~\ref{4dems}).
These results suggest that the overall shape of the observed DEM can be very well reproduced with a
collection of hot loops ($T_{\text{max}} >$ 1 MK) that have significant expansion factors
($\gamma\sim$ 5 -10). However, the peak value of the DEM as predicted by the model is larger than
the observed value by a factor of about 100. Therefore, the loops must fill only a small fraction of
the coronal volume (filling factor $\sim 1$ \%).

\section{SUMMARY AND DISCUSSION}
Using the high temporal cadence observations from the HMI and AIA instruments on board SDO, we
studied the cases of emerging bipolar regions in the quiet Sun. High cadence data from AIA including
six EUV channels are re-sampled to 1 minute data to improve the signal-to-noise ratio, as well as,
to have a good temporal resolution. Further, \verb+xrt_dem_iterative2.pro+ is used to construct DEMs
near loop top in a $6'' \times 6''$ pixel region (Section~\ref{sec:obsvr}, Figure~\ref{4dems}). From
these DEMs, we get the temporal evolution of emission weighted temperature with
Equation~\eqref{emwt}. 

Integrated unsigned magnetic flux derived from the HMI observations (10$^{19}$ $-$ 10$^{20}$~Mx) is
compared with the temperature of the loop, for each example (Figure~\ref{4examp}). There is no clear
relation between the two quantities, suggesting that, for these small emerging bipoles, the
surrounding regions in photosphere and higher atmosphere play an important role in the loop
evolution. 

To estimate the energetics involved in the formation of these loops we use a hydrodynamic loop
model (EBTEL) to simulate the DEMs and emission weighted temperatures. We assume that a loop is a
bundle of one hundred strands, each having a length of about 18 Mm, and a uniform radius of 0.1 Mm.
Furthermore, each strand is randomly heated and the average effect describes the properties of the
observed loop. To this end, we tested three simple heating events with varied heating frequencies as
described in Sections~\ref{sec:case1},~\ref{sec:case2}, ~\ref{sec:case3a},
and~\ref{sec:case3b}. The average heating input in our study ($\approx$ 10$^{6}$
ergs cm$^{-2}$ s$^{-1}$) is in close agreement with the approximate energy
losses observed in the quiet Sun. The $3\sigma$ values of the fluctuations in $T_{DEM}$ (MK)
are about 0.25, 0.16, 0.55, and 0.24 for the observations, cases 1, 2, and 3(a) respectively. 

In case 3,  we tested the following sub-cases: (a) 20\% of the strands are heated to high heating
values all the time, and the remaining strands are heated to low heating values
(Section~\ref{sec:case3a}), (b) all strands are heated to high heating values 20\% of their time
(Section~\ref{sec:case3a}), (c) similar to case 3(a), but for expanding strands to account for the
adiabatic cooling effects (Section~\ref{sec:case3b}). It is shown that the cases 3(a), and 3(b) are
equivalent and match the observations fairly well. This suggests that there may be a range of
heating events operating in the loops at a given time. In the cases 2 and 3(b), though there are
only a few large heating events, the essential difference between the two cases is that, unlike in
case 2, the duration of a single high heating phase in case 3(b) itself is longer compared to
typical plasma cooling time (making case 3(b) statistically a steady heating model).
This allows case 3(b) to find a DEM peak at higher temperatures. The model described in case 3 is
the best model we could obtained with in the scope of the present work. Mixing low and medium
heating at various proportions with different average heating rates show discrepancy, and do not fit
observations completely. These results are based on the assumption that the emitting plasma
has coronal origin.

Alternatively, we also argued that to include TR emission in the model, it is important to consider
an expansion of the loop from TR to corona. In this scenario, a steady heating model for loops with
loop apex temperature $> 10^6$ K can well reproduce the observed DEMs, assuming $\sim$ 1\% plasma
filling factor.

Reliability of AIA DEMs is a matter of debate.~\citet{2010A&A...521A..21O} studied the contribution
of spectral lines and continuum emission to the AIA EUV channels using CHIANTI atomic database. They
emphasize that the contribution of particular spectral lines and continuum emission can affect the
interpretation of the observed features, when AIA channels are used to observe regions other than
those for which the channels were designed.~\citet{2011A&A...535A..46D} compared AIA DEMs with the
\textit{Hinode}/EIS observations of active regions. They found discrepancies between the derived
DEMs. This is mainly due to the multi-thermal nature of AIA response curves, which have
contributions from cooler components. The cooler emission below 6.0 (in log $T$), seen in our
observations could be due to the double-peaked nature of AIA responses as suggested
by~\cite{2011A&A...535A..46D}. Empirically modified filter response curves for AIA are derived to
address some of these issues however, the role of this possible contamination in a already cool loop
(like the one originating from a small bipole in the quiet Sun), as compared to the warm loops in
the hotter active region has to be further examined. 

The models presented in this work assume that the strands are heated uniformly over their entire
length. Alternatively, the strand can be heated in a non-uniform manner with localized and
concentrated heat sources. If the heating is concentrated at the loop footpoints, this may lead to
the loss of equilibrium in the energy balancing terms, as the radiative losses in the coronal
section dominate the downward conductive flux. This will trigger the runaway cooling due to strong
radiative losses and a condensation is formed in the coronal loops~\citep[for
example][]{1980A&A....87..126H,2004A&A...424..289M}. This is a well studied phenomenon in the
formation of solar prominences~\citep{1991ApJ...378..372A,1999ApJ...512..985A}. Recently, based upon
the observed properties of the hot, and warm loops in active regions, ~\citet{2010ApJ...714.1239K}
have argued that the high concentration of heating low in the corona, and the steady or quasi-steady
heating models (leading to thermal nonequilibrium) can be ruled out. However,
~\citet{2012A&A...537A.152P} claim that a steady supply of energy is required even in the events of
condensation in the corona, to keep the coronal pressure. They also suggested that thermal
non-equilibrium can be a valuable tool in investigating the plasma dynamics and heat input in the
regions where condensation forms. 

The studies on the role and importance of the thermal non-equilibrium in the formation of
condensation in the short quiet Sun loops are not extensive.
~\citet{2003A&A...411..605M},~\citet{2004A&A...424..289M} discussed in detail the numerical
simulations of condensation and catastrophic cooling of short TR 10 Mm loops, and
longer 100 Mm coronal loops, respectively. They considered heating that has exponential height
dependence along the loop, and further suggested that the catastrophic cooling is initiated by the
loss of equilibrium at the loop apex due to concentration of heating at the footpoints, but not due
to a drastic decrease of the total loop heating. 

Note that the strands in a loop may interact in a very complex manner, and their response to the
condensation is the key objective to be addressed. Further work is necessary to get a better picture
of the nature and location of the heating, observational signatures of condensation, and finally the
role of magnetic field in this whole process. A complete set of answers for these questions is still
elusive and we need more observational constraints to narrow down the possibilities.

\acknowledgments 
The authors thank the referee for many comments and suggestions that helped in improving the
presentation of the manuscript. LPC. is a $2011-2013$ SAO Pre-Doctoral Fellow at the
Harvard-Smithsonian Center for Astrophysics. Funding for LPC and EED is provided by NASA contract
NNM07AB07C. Funding from the Indo-Austria exchange program (INT/AUA/BMWF/P-11/2011) is acknowledged.
LPC thanks Jim Klimchuk, Steve Saar, and Mark Weber for many useful suggestions and discussions.
Courtesy of NASA/SDO and the AIA and HMI science teams. This research has made use of NASA's
Astrophysics Data System.
\clearpage
\appendix
\section{EBTEL and Loops That Expand with Time}
\label{sec:appendixA}
The standard version of EBTEL assumes a symmetric loop with constant loop length and, uniform
cross-section.
The model is based on the 1D time-dependent energy conservation equation
\begin{equation}
\frac{\partial E}{\partial t}= -\frac{\partial}{\partial s}v(E+P) - \frac{\partial F_c}{\partial s}
+ Q  - n^2\Lambda(T) \label{ebtel1},
 \end{equation}
where $s$ is a spatial coordinate along the magnetic field; $E = \frac{3}{2}P + \frac{1}{2}\rho v^2$
is the total energy density; $n, T, P, \text{and}~v$ are the electron number density, temperature,
total pressure, and plasma bulk velocity, respectively; $F_c$ is the heat flux; $Q$ is volumetric
heating rate; and $\Lambda(T)$ is the radiative loss function for optically thin plasma. It is
assumed that the velocity and heat flux both vanish at the loop apex due to symmetry. Also, the flow
velocity is subsonic, and gravity is neglected in the energy equation. Integrating the above
equation over the coronal ($L$), and TR ($l$) lengths with the above assumption, we
get
\begin{eqnarray}
\frac{3}{2}L\frac{\partial\bar P}{\partial t} &\approx& \frac{5}{2}P_0v_0+F_0+L\bar Q-\mathcal{R}_c,\\
\frac{3}{2}l\frac{\partial\bar P_{tr}}{\partial t} &\approx&-\frac{5}{2}P_0v_0-F_0+l\bar Q_{tr}-\mathcal{R}_{tr},
\end{eqnarray}
where overbar denotes the spatial averages of the quantities over respective sections of the loop,
and subscript 0 denotes the values at the base of the corona; $\mathcal{R}_c,
\text{and}~\mathcal{R}_{tr}$ are the coronal, and the TR radiative loss rates
respectively. Neglecting the terms involving $l$ (for a thin TR), and together with
ideal gas law\footnote{$P=2nkT$, where $k$ is the Boltzmann's constant.}, $\bar P$ and $\bar n$ can
be approximated\footnote{simple volumetric averaging yields similar results.} with
\begin{eqnarray}
\frac{d\bar P}{dt} &\approx& \frac{2}{3}\left[\bar Q-\frac{1}{L}(\mathcal{R}_c + \mathcal{R}_{tr})\right], \label{eq:pbar}\\
\frac{d\bar n}{dt} &\approx& -\frac{1}{5kLT_0}\left(F_0+\mathcal{R}_{tr}\right)\label{eq:nbar}.
\end{eqnarray}
For a given heating rate $\bar Q(t)$, the EBTEL model returns $\bar P, \bar n, \text{and}~\bar T$, with other useful quantities.

For a uniformly expanding strand of length $L(t)$ and radius $R(t)$ adiabatically, the above
equations are modified by adding a term $-\gamma\bar P\xi(t)$ on the right hand side of
Equation~\eqref{eq:pbar}, and $-\bar n\xi(t)$ in Equation~\eqref{eq:nbar}, where $\gamma=5/3$ is the
ratio of specific heats, and
\begin{equation}
 \xi(t)=\frac{1}{L}\frac{dL}{dt}+\frac{2}{R}\frac{dR}{dt}\label{eq:xi}. 
\end{equation}
The strand pressure, density, and temperature are modeled accordingly. The time varying length and
radius of a single strand are shown as thin solid and dashed curves, respectively, in
Figure~\ref{volume}. Note that $R(t)$ explicitly enters the scheme only through
Equation~\eqref{eq:xi}, and everywhere else, it is absorbed due to volumetric averaging.

\section{Model for Loops that Expand with Height}
\label{sec:appendixB}

In this section we describe a loop model for the case that the
cross-sectional area $A$ varies along the loop. For simplicity the
area $A(T)$ is considered to be a function of temperature:
\begin{equation}
A(T) = \exp \left\{ \ln \gamma \left[ \frac{z(T)} {z(T_{\rm max})}
- 1 \right] \right\} , \label{area} 
\end{equation}
where $T_{\rm max}$ is the maximum temperature at loop top, and $z(T)$
is a monotonically increasing function, starting with $z \approx 0$ at
the base of the TR. We use $z(T) = y + \sqrt{1+y^2}$ with $y =
(x-x_0)/x_1$ and $x = \log_{10} T$. The constants $x_0$ and $x_1$ are
set to 6.0 and 0.2, respectively, so that most of the area change
occurs near a temperature of 1 MK. Similarly, the volumetric heating
rate is
\begin{equation}
Q(T) = Q_{\rm max} \left( \frac{T} {T_{\rm max}} \right)^m , \label{heat}
\end{equation}
where $Q_{\rm max}$ is the heating rate at the loop top, and $m$ is an
exponent (for the models presented here we set $m=0$). The loop is
assumed to be symmetric, and heating is assumed to be steady in time.
We solve the following energy balance equation: 
\begin{equation}
\frac{\partial}{\partial s} \left( A F_c \right) =
A(T) \left[ Q(T)  - n^2 \Lambda(T) \right] , \label{steady}
\end{equation}
where $s$ is a spatial coordinate along the magnetic field,
$F_c (s) \equiv - \kappa_0 T^{5/2} \partial T / \partial s$ is the
conductive heat flux, $n(s)$ is the electron density, and $\Lambda(T)$
is the radiative loss function, which is taken from 
\citet{2008ApJ...682.1351K}. Multiplying equation (\ref{steady}) by
$A F_c$ and integrating over position along the loop, we obtain:
\begin{equation}
\onehalf A^2 F_c^2 = \kappa_0 Q_{\rm max} [ f E_1 (T) - E_2 (T) ] ,
\end{equation}
where
\begin{eqnarray}
E_1 (T) & = & \int_{T_{\rm base}}^T A^2 (T) \Lambda (T) T^{1/2} dT , \\
E_2 (T) & = & \int_{T_{\rm base}}^T A^2 (T) ( T/T_{\rm max} )^m T^{5/2} dT .
\end{eqnarray}
Here $T_{\rm base}$ is the temperature at the base of the TR
($T_{\rm base} = 10^4$ K), and we assume $F_c = 0$ at the base.
The factor $f$ is given by
\begin{equation}
f \equiv \frac{P^2} {4 k^2 Q_{\rm max}} = \frac{E_2(T_{\rm max})}
{E_1(T_{\rm max})} , \label{fff}
\end{equation}
where $P = 2nkT$ is the plasma pressure (a constant), and the last
equality in (\ref{fff}) follows from the requirement that $F_c = 0$ at
the loop top. Then the loop half-length $L$ is given by 
\begin{equation}
L = \int_{s_{\rm base}}^{s_{\rm max}} ds = 
\left( \frac{\kappa_0} {2 Q_{\rm max}} \right)^{1/2}
\int_{T_{\rm base}}^{T_{\rm max}} 
\frac{A(T) T^{5/2} dT} {\sqrt{ f E_1(T) - E_2(T) }} .
\end{equation}
For a given peak temperature $T_{\rm max}$ and half-length $L$, we can
compute the heating rate $Q_{\rm max}$, pressure $P$, heat flux
$F_c (T)$, and density $n(T)$. Then the DEM is given by $\varphi(T) =
n^2(T) A(T) (\partial T/ \partial s)^{-1}$. Since the area factor is
normalized such that $A(T_{\rm max}) = 1$, this DEM($T$) does not
include the effects of a possible filling factor of the coronal
loops.



\end{document}